\documentclass[twocolumn,aps,10pt,prl,showpacs,showkeys,]{revtex4}
\usepackage[dvips]{graphicx}

\begin{document}
\title{LaCoO$_{3}$ - from first principles}
\author{R. J. Radwanski}
\affiliation{Center of Solid State Physics, S$^{nt}$Filip 5,
31-150 Krakow, Poland\\
Institute of Physics, Pedagogical University, 30-084 Krakow,
Poland}
\author{Z. Ropka}
\affiliation{Center of Solid State Physics, S$^{nt}$Filip 5,
31-150 Krakow, Poland} \homepage{http://www.css-physics.edu.pl}
\email{sfradwan@cyf-kr.edu.pl}

\begin{abstract}
We have performed calculations of the electronic structure of
LaCoO$_{3}$ from first principles, assuming the atomistic
construction of matter and the electrostatic origin of the
crystal-field splitting. In our atomic-like approach QUASST the
{\it d} electrons of the Co$^{3+}$ ion in LaCoO$_{3}$ form the
highly-correlated atomic-like system 3$d^{6}$ with the singlet
ground state $^{1}A_{1}$ (an octahedral subterm from the $^{1}I$
term) and the excited octahedral subterm $^{5}T_{2g}$ of the
$^{5}D$ term. In the spin-orbital space, being physically
adequate, this high-spin state is Jahn-Teller active.  We take
the ESR experiment of Noguchi {\em et~al.}, Phys. Rev. B {\bf
66}, 094404 (2002), as confirmation of the existence of the
discrete electronic structure for 3$d$ electron states in
LaCoO$_{3}$ in the meV scale postulated in QUASST.

\end{abstract}
\maketitle \vspace {-0.3 cm}

LaCoO$_{3}$ exhibiting non-magnetic ground state and the
significant violation of the Curie-Weiss law at low temperatures
\cite{1} often discussed in terms of successive changes of spin
states (low- (LS), intermediate- (IS) and high-spin (HS) states)
with the increasing temperature attracts much attention in recent
50 years. Despite of large activity, both theoretical and
experimental, there is still enormous chaos in theoretical
understanding of its properties. We would expect that the problem
of LaCoO$_{3}$, of the origin of the excited state in particular,
has been clarified in 2003 in our paper \cite {2}, making use of
experimental results of Noguchi {\em et~al.} \cite{3}, but
recently has appeared a paper in Phys. Rev. Lett. of Phelan {\em
et~al.} \cite {4} with a claim that i) the excited state in
LaCoO$_{3}$ is the intermediate-spin S=1 state of a
$t_{2g}^{5}e_{g}^{1}$ configuration.  Moreover, they have claim
that ii) the HS state $t_{2g}^{4}e_{g}^{2}$ (S=2), in contrary to
the IS state, is not Jahn-Teller active and that iii) the
$t_{2g}$-$e_{g}$ splitting in LaCoO$_{3}$ is small. The
appearance of the paper of Phelan {\em et~al.} in Phys. Rev.
Lett. is a direct motivation for this paper. The aim of this
paper is to clarify above mentioned erroneous claims and to
present a consistent understanding of LaCoO$_{3}$ within the
localized atomistic paradigm. We understand atomistic ionic
paradigm as the most natural and physically adequate approach to
transition-atom compounds.

The IS state as the first excited state has been introduced to
the LaCoO$_{3}$ problem in year of 1996 by band calculations of
Korotin {\em et~al.} \cite{5} as an {\bf opposite} view to the
ionic view being a base for the Tanabe-Sugano diagrams known from
years of 1954 and 1970 \cite{6}. The Tanabe-Sugano diagram for
the 3$d^{6}$ configuration, Fig. 1, has yielded, accepting the
relatively strong crystal field Dq/B$>$2, the excited state to be
the HS $t_{2g}^{4}e_{g}^{2}$ state and this view was the base for
a model of Goodenough \cite {1}. The IS-state concept of Korotin
{\em et~al.} became highly popular with hundreds of citations
\cite {7,8,9,10,11,12}. In the band calculations of Korotin {\em
et~al.} the IS state becomes the first excited state as an effect
of the especially strong $d-p$ hybridization. However, we claim
that if at present, in year of 2006, one wants to still claim
that the IS state is an excited state has to present a
quantitative band-based or hybridization-based interpretation of
the Noguchi {\em et~al.} experiment.

In a situation of the dominant band view in LaCoO$_{3}$ problem,
in general in 3d oxides \cite {5,7,8,9,10,11,12}, and of
continuous claiming of the IS state as the first excited state we
feel necessary to present our understanding of properties of the
magnetism and electronic structure of LaCoO$_{3}$ in the
localized atomistic paradigm. This our understanding is based on
well-known physical concepts like the crystal-field (CEF),
spin-orbit (s-o) coupling, local distortions and other terms known
from the ionic language. We present this view being aware that the
ionic picture and crystal-field considerations are at present
treated as the "old-fashioned" and contemptuous physics in times
of wide spreading omnipotent band theories of different versions
LDA, LSDA, LDA+U, LDA+GGA, DMFT and many, many others. We gave a
name of QUASST for our approach to a solid containing
transition-metal atoms from Quantum Atomistic Solid State Theory
pointing out that the physically adequate description of
properties of a 3d/4f/5f solid the best is to start from analysis
of the electronic structure of constituting atoms.

\begin{figure}\vspace {+0.5 cm}
\begin{center}
\includegraphics[width = 10.5 cm]{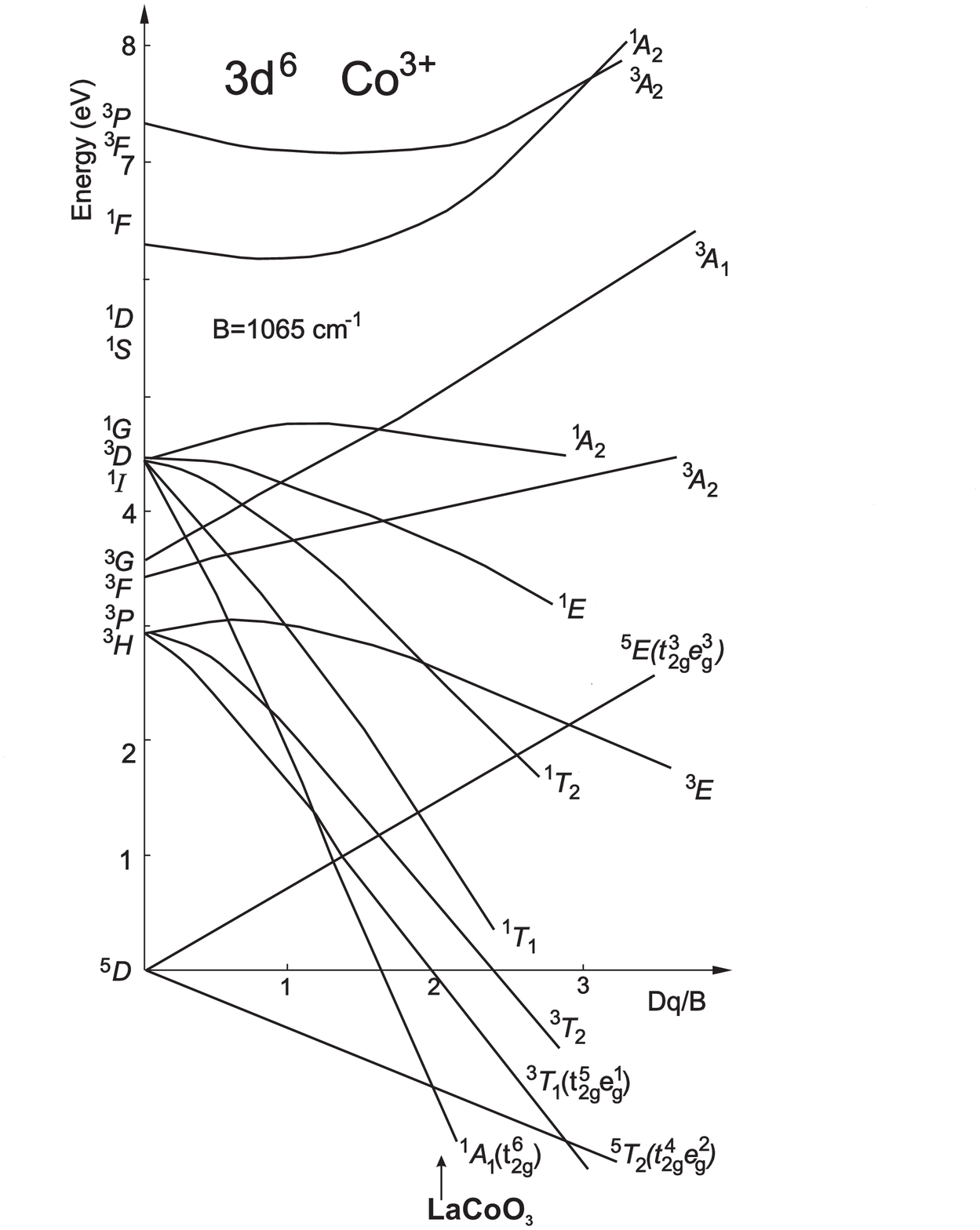}
\end{center} \vspace {-0.1 cm}
\caption{A modified Tanabe-Sugano diagram for the Co$^{3+}$ ion
(3$d^{6}$ configuration) showing the effect of the octahedral
crystal field on the electronic terms of the free Co$^{3+}$-ion.
The electronic structure of cubic subterms, corresponding to
$Dq$/$B$ =2.025, relevant to LaCoO$_{3}$ is marked by an arrow.
\vspace {-0.6 cm}}
\end{figure}

Schematic steps of the QUASST approach to LaCoO$_{3}$ can be
written as:

1. we accept the atomistic structure of matter, i.e. atoms
preserve much of their atomic properties becoming the full part
of a solid like, for instance, in NiO or LaCoO$_{3}$. In
LaCoO$_{3}$ during the formation of compound there occurs an
electron transfer of three electrons from La and Co to O with the
simultaneous formation of the perovskite lattice with the ionic
charge distribution La$^{3+}$Co$^{3+}$O$_{3}^{2-}$,

2. all magnetism of LaCoO$_{3}$ is related to the Co$^{3+}$ ion
with six electrons outside close configuration $^{18}$Ar because
both O$^{2-}$ and La$^{3+}$ ions have closed shells,

3. six electrons of the Co$^{3+}$ ion form strongly-correlated
atomic-like system, 3d$^{6}$; its electronic structure is known
from the atomic physics,

4. from the perovskite structure of LaCoO$_{3}$ we learn that:

a) all Co sites are equivalent,

b) the local symmetry of the Co ion is octahedral,

c) the Co ion is surrounded by six nearest neighbours oxygens
forming almost perfect octahedron,

d) there appears small off-octahedral distortion with a lowering
temperature.

 5. influence of the octahedral crystal field on the free-ion electronic
terms,

6. influence of the intra-atomic spin-orbit coupling on the local
electronic structure,

7. influence of the rhombohedral distortion, octahedral crystal
field on the local electronic structure,

8. magnetic interactions are not considered in LaCoO$_{3}$ because
LaCoO$_{3}$ up to lowest temperatures does not form
magnetically-ordered state,

9. having determined the local electronic structure with the
eigenfunctions we have the zero-temperature properties as well as
the free energy $F$(T),

10. the zero-temperature properties are related to the properties
of the local ground state; the free energy $F$(T) enables
calculations of the whole thermodynamics. See, for instance, our
description of FeBr$_{2}$, an exemplary of 3$d$ compounds.

Ad. 1. An assumption about the atomistic construction of matter
seems to be obvious but we would like to point out that
band-structure calculations (LDA, LSDA, ..) disintegrate 3$d$
atoms completely starting from consideration of 3$d$ electrons as
independent electrons in the octahedral crystal field. According
to us these one-electron calculations has to reproduce, before
calculations of properties of a solid, the electronic structure of
the given ion. In our approach we simply accept the term
electronic structure known from the atomic physics as is clearly
written in p. 3 pointing out the importance of strong electron
correlations among $d$ electrons for the formation of the
intraatomic term structure.

Ad. 3. From the atomic physics we know that for the 3d$^{6}$
configuration of the Co$^{3+}$ ion 210 states are grouped in 16
terms: $^{5}D$ (25 states),  $^{3}H$ (33),  $^{3}G$ (27), $^{3}F$
(21), $^{3}F$ (21), $^{3}D$ (15), $^{3}P$ (9), $^{3}P$ (9),
$^{1}I$ (13), $^{1}G$ (9), $^{1}G$ (9), $^{1}F$ (7), $^{1}D$ (5),
$^{1}D$ (5), $^{1}S$ (1) and $^{1}S$ (1). These terms are shown
in Fig. 1 in the energy scale. The two Hund rules ground term of
the free Fe$^{2+}$ and Co$^{3+}$ ions is $^{5}D$, a number of
excited terms are at least 3 eV above. Experimental energies of
the free-ion terms have been tabulated in NIST \cite {14}. To
these $L-S$ coupling terms an effect of the $j-j$ coupling can be
added, if necessary.

Ad. 4. For more complex structures there will be a few 3$d$ sites,
each of them having own electronic structure.

Ad. 5. Influence of the octahedral crystal field on the free-ion
electronic terms has been calculated by Tanabe and Sugano in a
year of 1954 already \cite {6}. The splitting of electronic terms
to octahedral subterms in a function of the strength of the
octahedral CEF parameter Dq/B (B - intra-ionic Racah parameter
introduced for in order to account theoretically the arrangement
of the free-ion terms determines the energy scale) is known as
Tanabe-Sugano diagrams. A modified Tanabe-Sugano diagram for
3$d^{6}$ is presented in Fig. 1. A problem was and still is with
the evaluation of the value of Dq/B on this diagram for a given
compound. Numerous qualitative indications, starting already at
fifties of the XX century, have not been conclusive. In the
crystal-field theory parameter 10Dq ($\cong$120$\cdot$B$_{4}$) in
the simplest form is the multiplication of the octupolar charge
moment of the lattice A$_{4}$ and of the involved cation caused
by anisotropic charge distribution of the own incomplete shell.
Thus Dq (B$_{4}$) can be calculated from first principles
provided the octupolar charge moment, $\beta$ $
\langle{r^{4}}\rangle$ of the involved ion is known. $\beta$ is
the fourth-order Stevens coefficient.

The octahedral crystal field coefficient A$_{4}$ that is the
octupolar charge moment of all surrounding charges at the Co site
can be calculated from the point-charge model which is
first-principles elementary calculations. Taking the charge of
oxygen as -2e and the cation-oxygen distance of 192.5 pm \cite
{12,13} in LaCoO$_{3}$ we obtain a value of A$_{4}$ of +432
Ka$_{B}^{-4}$, a$_{B}$  is the Bohr radius. Taking for the
Co$^{3+}$ ion $\beta$ = +2/63 and  $ \langle{r_{d}^{4}}\rangle$
=2.342 a$_{B}^{4}$ \cite {15} we get B$_{4}$ = +32 K. This value
is eight times smaller than the recent evaluation of B$_{4}$ of
+260 K \cite {2}. In Ref. 2 we pointed out that despite this
difference the most important is that 1) these {\it ab initio}
calculations give the proper sign of the B$_{4}$ parameter
because it determines the ground state in the oxygen octahedron
and 2) the experimentally derived strength of crystal-field
interactions turns out to be much weaker than it was thought so
far in literature for justification of the strong crystal-field
approach in which one-electron approach becomes more physically
adequate. Here we would like to explain this 8-times difference
by attributing it to a large underestimation of
$\langle{r_{d}^{4}}\rangle$ =2.342 a$_{B}^{4}$ in Hartree-Fock
calculations from a year of 1965 \cite {15}. Good agreement is
obtained if $\langle{r_{d}^{4}}\rangle$ would be about 18
a$_{B}^{4}$. This later larger value is in agreement with recent
calculations. Korotin {\em et~al.} \cite {5} have used a value
for $r_{d}$ of 1.26 $\AA$ yielding approximately
$\langle{r_{d}^{4}}\rangle$ = 32.2 a$_{B}^{4}$. Recently Solovyev
\cite {16}, p. 5, has calculated for the Ti$^{3+}$ ion
$\langle{r_{d}^{2}}\rangle$ of 2.27 $\AA$$^{2}$ (=8.11
a$_{B}^{2}$), which approximately yields
$\langle{r_{d}^{4}}\rangle$ even of 65 a$_{B}^{4}$. Thus we think
that a value of 18 a$_{B}^{4}$ needed for the ionic
electrostatic-origin of the crystal-field splitting in LaCoO$_{3}$
is fully reasonable. Concluding, we think that too small values
obtained so far for the CEF parameters were largely due to too
small values taken for the octupolar moment of the
transition-metal atom, in particular for
$\langle{r_{d}^{4}}\rangle$.

Ad. 6. In the simplest form the spin-orbit Hamiltonian takes the
form $H_{s-o}$ = $\lambda_{s-o}$L$\cdot$S for the given term or
with a parameter $\xi$ for the whole configuration. Values for
$\lambda_{s-o}$ for different configurations of free 3$d$ ions
have been collected in text-books \cite {17}. Of course, the s-o
coupling in a compound can be different from the free-ion value.

Ad. 7. LaCoO$_{3}$ develops below 1610 K a slight rhombohedral
(trigonal) distortion reaching an angle 60.79 degrees at 4 K
\cite {1,5}. It means that the local oxygen octahedron is
compressed along the diagonal. We have calculated that for such
distortion the parameter B$_{2}^{0}$ is positive like it was
found in Ref. \cite {2}. The distortion is small and causes a
slight splitting of the lowest quasi-triplet with $D$ = 4.9
cm$^{-1}$ as has been revealed experimentally by Noguchi {\em
et~al.} and described by us in Ref. \cite {2} by a trigonal
parameter $B_{2}^{0}$ of +7.2 K.

Ad. 8. The formation of the magnetic state we have described in
numerous compounds - let mention exemplary 4f/3d/5f compounds
ErNi$_{5}$, FeBr$_{2}$, NiO, CoO, UPd$_{2}$Al$_{3}$ and
UGa$_{2}$, results of which have been published starting from
1992. In all these cases the magnetic energy is much smaller than
the overall CEF splitting. They are both ionic (FeBr$_{2}$, NiO,
CoO) and intermetallic (ErNi$_{5}$, UPd$_{2}$Al$_{3}$, UGa$_{2}$)
compounds \cite {18}.

We appreciate very much Refs \cite {4,5} but in contrary to a
claim of Ref. 4 that ii) the high-spin HS state
$t_{2g}^{4}e_{g}^{2}$ (S=2) is not Jahn-Teller active in contrary
to the IS state we argue that the HS state (S=2) is Jahn-Teller
active, see Fig. 2 of Ref. \cite {19}, Fig. 1 of Ref. \cite {20}
and Fig. 6 of Ref. \cite {2}. We have claimed that the
Jahn-Teller effect has to be considered in the spin-orbital
space, but not in the orbital space only \cite {21}. In contrary
to a claim of Ref. 4 that iii) the $t_{2g}$-$e_{g}$ splitting in
LaCoO$_{3}$ is small we argue that the crystal-field is
relatively strong, the best described as indermediate. In
contrary to a result of Ref. 5 revealing the spin polarization in
the IS and HS state in the paramagnetic state we argue that in
the reality there is no spin polarization. We do not agree with a
claim that "the crystal-field energy is determined by the $3d-2p$
hopping parameters, ..." - in our understanding the crystal-field
splitting results from conventional electrostatic interactions of
the multipolar character, it means it has the same origin as
Stark effect known in the atomic physics. Finally we note
substantial difference of our approach with that of Ref. \cite
{5} with respect to the actual $d$-shell occupation. Instead of
the six-electron configuration in the ionic model considered by
us Ref. \cite {5} yields 7.3 $d$ electrons in average. It means
that Korotin {\em et~al.} calculations provide much smaller
electron transfer and realization of the smaller valency.

We would like to note that all of the used by us parameters
(dominant octahedral CEF parameter B$_{4}$, the spin-orbit
coupling $\lambda_{s-o}$, lattice distortions) have clear physical
meaning and can be calculated from first-principles. The most
important assumption is the existence of very strong correlations
among 3$d$ electrons preserving the atomistic ionic integrity of
the Co$^{3+}$ ion also in the solid when this cation becomes the
full part of a solid in LaCoO$_{3}$. The electronic structure of
the transition-metal atom predominantly determines the
macroscopic properties of the whole compound containing
transition-metal 3d/4f/5f atoms. With pleasure we notice the good
reproduction of experimental magnetic susceptibility and heat
capacity with our energy level structure \cite {2} by recent
paper of Kyomen {\em et~al.} \cite {22}.

In {\bf conclusions}, we have calculated from first-principles
the low-energy electronic structure of LaCoO$_{3}$ which
determines its macroscopic properties. The electronic structure
originates from the Co$^{3+}$ ions taking into account the
calculated octahedral CEF interactions, the intra-atomic
spin-orbit coupling and a relatively weak trigonal distortion and
assuming the atomistic integrity in LaCoO$_{3}$ of the Co$^{3+}$
ion with the 3$d^{6}$ configuration.

Our calculations explain the non-magnetic ground state as related
to the ionic $^{1}A_{1}$ ($^{1}I$) subterm, the first-excited
state as related to the HS $^{5}T_{2g}$ ($^{5}D$) and insulating
ground state. Our calculations confirm the physical adequacy of
the Tanabe-Sugano diagrams and we quantify the strength of the
octahedral CEF parameter of the Co$^{3+}$ ion in LaCoO$_{3}$ as
B$_{4}$ = +260 K ($Dq$/$B$=2.025 on the Tanabe-Sugano diagram).

The crystal-field interactions are relatively strong in
LaCoO$_{3}$ but not so strong to destroy the ionic integrity of
the 3$d$-electrons as is postulated in QUASST \cite {17}. Our
model can be classified as an ionic model but we have got a
number of nontrivial results not obtained so far in the ionic
model like the excited HS state magnetic moment of 0 and
$\pm$3.35 $\mu_{B}$ \cite {2} instead of 4 $\mu_{B}$ expected for
$S$=2 state. We are fully aware of many oversimplifications of our
approach but we strongly believe that it is a physically adequate
model to be further developed. Our long-lasting studies as well
as growing number of more and more sophisticated experiments
indicate that it is the highest time to ''unquench'' the orbital
moment in the solid-state physics for description of the
magnetism and the electronic structure of 3$d$-atom containing
compounds.

-----

{\bf Dedicated} to all researchers of LaCoO$_{3}$ - in our
understandings we have benefited from all conducting integral
research, both theoretical and experimental.

\end{document}